# Statistical parameters for assessing environmental model performance related to sample size: Case study in ocean color remote sensing

Weining Zhu

Department of Marine Informatics, Ocean College, Zhejiang University

**Abstract**
Environmental model performances need to be assessed using some statistical parameters, such as mean absolute error (MAE) and root mean square error (RMSE). The advantages and disadvantages of these parameters are still in controversial. The purpose of this study is to introduce a statistical parameter, type A uncertainty ($U_A$), into model performance evaluations. We particularly focus on the relations between sample sizes and three evaluation parameters, and tested a few ocean color remote sensing algorithms and datasets. The results indicate that RMSE, MAE and $U_A$ all vary with the sample size *n* but present different trends. Based on our tested results and theoretical analysis, we therefore conclude that $U_A$ is better than RMSE and MAE to express model uncertainty, because its downward trends indicate that the more samples we take, the less uncertainty we get. RMSE and MAE are good parameters for assessing model accuracy rather than uncertainty.

## 1. Introduction

In quantitative remote sensing of environment, many classification, correction and inversion models or algorithms need to be validated by in-situ collected samples, and then compare them with model's predictions, by calculating some well-known statistical parameters such as the mean absolute error (MAE), standard deviation (SD), root mean square error (RMSE), and bias. With respect to these parameters, we then can evaluate performances of these models/algorithms, in terms of their errors, accuracy, precision, effectiveness, robustness, etc.[1,2] These statistical parameters, however, are often simply adopted in remote sensing modelling/validation without deep explorations. If we check the newest articles about metrology, geophysics, and environmental science, these model evaluation parameters, even the error evaluation methodology, are still in controversy [3-5]. For example, the most commonly used two parameters: RMSE and MAE, have not been yet agreed that which one is more appropriate for model evaluation. Since the early eighties of the last century, some scholars, such as Prof. Willmott at University of Delaware, have paid attention to model assessments in earth, geographic information, and environmental science [3, 6-10]. They found the magnitude of RMSE being controlled by three factors: (1) MAE, (2) the distribution of error, and (3) sample size $n$ [3,9]. They have exemplified some models bearing the same MAEs but corresponding to different RMSEs due to different error distributions. They therefore concluded that MAE is a more accurate parameter than RMSE for quantifying the magnitude of model errors, and then suggest using MAE rather than RMSE to evaluate model performance. In addition, Willmott et al. [3] also claimed that another drawback of RMSE is that it does not satisfy the triangle inequality of space metric.

If above claims of Willmott et al. are correct, then it implies that many RMSE-based model evaluations and validations in remote sensing science may not be correct.

Responding to the claims of Willmott et al., Chai and Draxler argued that RMSE is more appropriate to present model performance than MAE when the error distribution is expected to be Gaussian (or normal) [4]. In addition, they also proved that RMSE satisfies the triangle inequality required for a generic metric. They believe that for evaluating model performance, any single statistical parameter has limitations – it only demonstrates one face of a model. In order to provide a comprehensive assessment of model performance, they suggest using multiple parameters together.

Basically, we agree with the arguments of Chai and Draxler. We have noticed that MAE and RMSE were both reported in some remote sensing studies involving model/algorithm developing, evaluation, and comparison.

Besides RMSE and MAE, sample size $n$ also plays an important role in field samplings as well as modelling and validations, but relations between $n$ and RMSE or MAE have not been fully investigated in remote sensing model evaluations. One generally does not care about $n$, provided it satisfies a statistically significant amount. For instance, many ocean color remote sensing studies only report how many water and spectral samples were taken at their study sites, how many samples were used for modelling/calibration, and how many samples were used for validation. MAE and/or RMSE were then used to assess model performance but their sample sizes may range from couples, dozens to hundreds, thousands, and even millions. Thus here arises the question: how to evaluate and compare models established on different sample sizes? For example, a chlorophyll remote sensing algorithm A1 was modelled by 30 samples and validated by 15 samples, and evaluation result is $RMSE_1 = 0.12$ (mg/L). At the same site, the second algorithm A2 was modelled by 2,000 samples and validated by 1,000 samples, and got $RMSE_2 = 0.23$. We may ask: which algorithm is better? In terms of RMSE, obviously A1 is better than A2 since $RMSE_1 < RMSE_2$. But if considering the sample sizes, we may trust A2 more than A1 because A2 was modelled by a thousand samples which is larger than the size of A1. Note that an alternative approach to compare two models is using the same dataset (with the same sample size), but this approach cannot always work because sometimes model validations are subject to sample conditions and model requirements. For example, if chlorophyll algorithm A1 uses blue-green band ratio, but A2 uses red-NIR ratio or more multispectral or hyperspectral bands, then A1 and A2 are unable to be compared using the same samples if the wavelengths required for A2 are not available in the samples.

A recent study published in '*Nature Reviews Neuroscience*' warned that the small sample sizes undermine the reliability of neuroscience [11]. Similarly, the same undermining may occur in environmental science since its research objects, such as earth surface, ecosystems, greenhouse gas cycles, human-nature interactions, and urban land-use dynamics, are possibly more complicated than nervous systems, making the reliability of environmental models based on small sample sizes to be questionable. To make an environmental model more reliable, within the allowable costs, we usually desire samples as many as possible (suppose there are no serious mistakes in sampling and measuring).

The objective of this study is exploring the impact of sample size *n* on model evaluation parameters (MAE and RMSE), and introducing a new parameter, 'type A' uncertainty ($U_A$), which is more related to sample size and originally proposed in metrology. $U_A$ comes from '*The Guide to the expression of Uncertainty in Measurement* (GUM)', first published in 1993 by ISO, jointly issued in 2008 by seven international organizations BIPM, OIML, ISO, IEC, IUPAC, IUPAP, IFCC, later approved by the International Laboratory Accreditation Cooperation Organization (ILAC), and now is effective for all nations [12]. Although $U_A$ has been already used to evaluate the uncertainty of measurements, we found it can be used to evaluate the uncertainty of models, and bears some advantages over MAE and RMSE. So far $U_A$ is barely used in environment and remote sensing studies [13]. In ocean color remote sensing, most of models/algorithms are still evaluated by MAE, RMSE, and other traditional statistical parameters [14,15].

In this study, we tested several ocean color datasets, models/algorithms, and some simulated error datasets with different distributions, demonstrated how statistical evaluation parameters (MAE, RMSE, and $U_A$) varied with sample size *n*, and concluded that compared to MAE and RMSE, $U_A$ is more associated to sample size and hence a more proper parameter to assess model performance regarding its uncertainty and reliability. In addition, theoretical analyses of relations between MAE/RMSE/$U_A$ and sample size also verified our conclusions.

## 2. Material and methods
### 2.1 Data sets

Two ocean color open datasets were used in this study, both provided by the International Ocean Color Coordinating Group (IOCCG): (1) IOCCG in-situ dataset, a subset of the NASA bio-optical marine algorithm dataset (NOMAD) [16], which is a publicly available, global, and high quality in-situ bio-optical data set for ocean color algorithm development and satellite data product validations. The in-situ dataset contains 656 observations. (2) IOCCG synthetic dataset [17], a dataset also served for ocean color algorithm testing. The synthetic dataset contains 500 samples which are simulated by using the well-known radiative transfer software Hydrolight. The two datasets both contain several inherent and apparent optical properties (IOPs and AOPs) for remote sensing modelling and validation. These IOPs and AOPs include absorption and back-scattering coefficients of water, concentration or absorption coefficients of chlorophyll-a, colored dissolved organic matter, and suspended particulate matter, remote sensing reflectance ($R_{rs}$) in visible wavelengths, etc.

For each data set, a bootstrap method was used to create testing data, that is, many subsets were randomly resampled with sample sizes from $n = 1$ to the whole samples ($n = 500$ or $656$) with 1 interval (i.e., $n = 1, 2, …, 500$, respectively). The above processing was repeated $m$ times to minimize the possible uncertainty made by random selections. For example, if $m = 400$, then there are 400 subsets with $n = 1$, and 400 subsets with $n = 2$, and so on, and therefore we may have a very large number of subset families from each dataset.

Moreover, in order to show the ideal cases of error distributions, the software MATLAB® was used to directly generate random numbers as errors, forming four error datasets with different error distributions: (1) normal, (2) exponential, (3) lognormal, and (4) uniform. Each error set contains 10,000 data. The bootstrapping of these error datasets is the same as the above resampling method of ocean color datasets.

**2.2 Algorithms**

Four representative ocean color remote sensing models were selected for retrieval of $a(440)$ (water's absorption coefficients at 440 nm) from the subsets of the IOCCG synthetic and in-situ datasets. These models are named by (1) L98: one-step spectral ratio algorithm [18], (2) MM01: spectral-ration algorithm with chlorophyll concentration as an intermediate link [19], (3) Carder: the empirical portion of Carder_MODIS algorithm [20], and (4) QAA (Lee et al, 2014): a quasi-analytical algorithm at version 6 [21]. The details of each algorithm could be referred to their original publications. Note that the first three models and QAA (version 4) have been officially tested and compared by using IOCCG synthetic and in-situ datasets and the results were shown in IOCCG Report Number 5, 2006 [17].

**2.3 Statistical parameters and validation methods**

The performance of each model was evaluated by MAE, RMSE, and $U_A$ which are calculated by

$$MAE = \frac{\sum_{i=1}^{n}|x_i^{mod}-x_i^{obs}|}{n} \qquad (1)$$

$$RMSE = \sqrt{\frac{\sum_{i=1}^{n}(x_i^{mod}-x_i^{obs})^2}{n}} \qquad (2)$$

$$U_A = \sqrt{\frac{\sum_{i=1}^{n}(x_i^{mod}-x_i^{obs})^2}{n(n-1)}} \qquad (3)$$

where $x_i^{mod}$ is the model's *i*-th prediction, $x_i^{obs}$ is the *i*-th observed ground truth correspondingly, $e_i = x_i^{mod} - x_i^{obs}$ is the error of the *i*-th sample.

The $U_A$ in Eq. (3) is the same one defined by GUM. From the perspective of metrology, it is also known as the standard deviation of the mean (or the standard error of the mean) or the variance of the mean. From Eq. (1) and (3), it is easy to know that $U_A$ can be expressed by $\frac{RMSE}{\sqrt{n-1}}$, approximately equaling $\frac{RMSE}{\sqrt{n}}$ if $n$ is large. However, according to Eq. (3), if $n = 1$, then we are unable to calculate $U_A$, so in order to make $U_A$ valid for any sample size, we slightly modified the definition of $U_A$. Because for large sample size $n$, $n(n-1) \approx n^2$, and hence Eq. (3) turns to

$$U_A = \frac{\sqrt{\sum_{i=1}^{n}(x_i^{mod}-x_i^{obs})^2}}{n} \qquad (4)$$

By this definition, $U_A$ is exactly equal to $\frac{RMSE}{\sqrt{n}}$.

For each sample size $n$, we have many different data subsets randomly selected, therefore the final RMSE, MAE, and $U_A$ are their mean (or median) values of these subsets. The difference between using mean and median will be discussed later.

In those simulated error datasets, each datum is a computer simulated $e_i = x_i^{mod} - x_i^{obs}$ and therefore $e_i$ can be directly substituted into formula (1) - (4) to calculate those parameters.

## 3. Results and discussion
### 3.1 Tests of ocean color remote sensing models and datasets

Testing results of four ocean color remote sensing models and two datasets are shown in Fig. 1 with subset number $m$ = 400. For IOCCG in-situ dataset, we only tested L98 and MM01 models because some wavebands required for Carder and QAA are not available in the dataset.

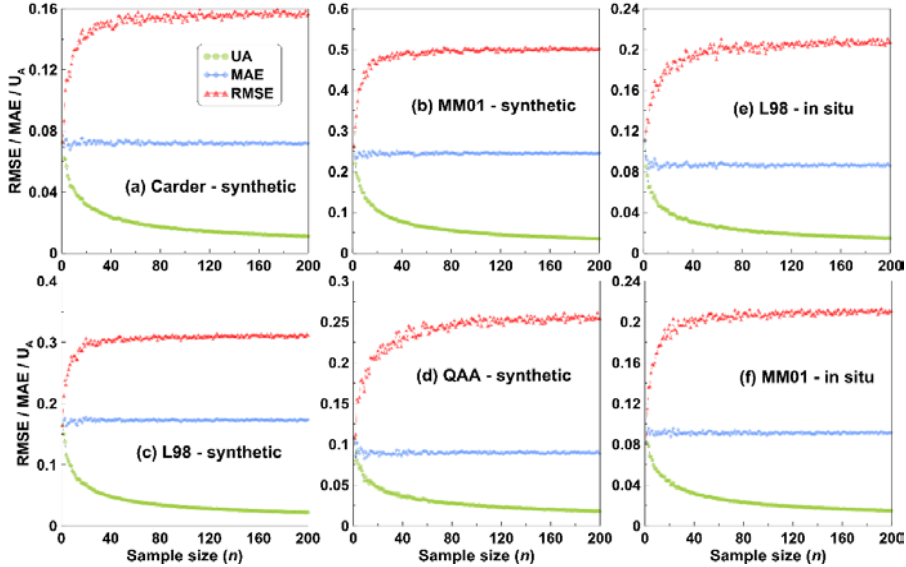

Figure 1. Statistical parameter trends vs. sample sizes calculated from four ocean color algorithms and two datasets.

The results in Fig. 1 illustrate that MAE, RMSE, and $U_A$ bear different trends as sample size increases. They all start from the same point ($n = 1$) where MAE = RMSE = $U_A$. After that, MAE, RMSE, and $U_A$ typically show the flat, upward, and downward trends, respectively. Data in Fig. 1 were only plotted to $n = 200$, because after this size, MAE, RMSE, and $U_A$ all smoothly approach to their respective expectations calculated from the whole samples, see Table 1. Table 1 shows the results of some statistical parameters calculated from the sample population. After about 20 ~ 50 samples, the upward trends of RMSE turn to be approximately flat.

Table 1. Statistical parameters for evaluating model performances, calculated from four ocean color algorithms and two IOCCG datasets. Note that $e_{max}$, $e_{min}$, and $e_{median}$ are all calculated from the absolute errors.

| Algorithm | Datasets | N | MAE($e_{mean}$) | $e_{max}$ | $e_{min}$ | $e_{median}$ | RMSE | $U_A$ |
|---|---|---|---|---|---|---|---|---|
| L98 | Synthetic | 500 | 0.173 | 1.756 | 2E-3 | 0.068 | 0.313 | 0.014 |
| MM01 | Synthetic | 500 | 0.245 | 2.521 | 5E-6 | 0.033 | 0.503 | 0.023 |
| Carder | Synthetic | 500 | 0.074 | 1.303 | 9E-6 | 0.018 | 0.160 | 0.007 |
| QAA | Synthetic | 500 | 0.090 | 2.741 | 7E-5 | 0.016 | 0.261 | 0.012 |
| L98 | In-situ | 656 | 0.107 | 2.176 | 2E-4 | 0.029 | 0.240 | 0.011 |
| MM01 | In-situ | 656 | 0.111 | 1.507 | 1E-4 | 0.027 | 0.242 | 0.011 |

Fig. 2 shows the error distributions of the six algorithm/dataset combinations, including the original, absolute, and squared errors. The distributions of errors of the tested algorithm/dataset are approximately with normal form (e.g., Carder/synthetic, L98/in-situ, and MM01/in-situ), lognormal (e.g., QAA/synthetic and L98/synthetic), or exponential (e.g., MM01/synthetic). Most of errors are in small values around the zero, and the rest errors are either symmetrically or asymmetrically distributed with respect to the zero. Asymmetrical distributions imply that there are significant biases, such as those of QAA, MM01 and L98 tested by the synthetic dataset. Due to the actions of making absolute and squared values, the absolute error and squared error are all with exponential-like distributions. The squared errors, which are used to calculate RMSE, illustrate even stretched tall-head and heavy-tailed distributions: because by squaring, the smaller becomes even smaller (e.g., $0.1^2 = 0.01$) while the larger becomes even larger (e.g., $10^2 = 100$).

### 3.2 Tests of simulated standard error distributions

The histograms of computer simulated errors with four distributions (normal, exponential, lognormal, and uniform) are shown in Fig. 3, and their results are shown in Fig. 4, in which the trends of MAE, RMSE, and $U_A$ vs. sample size $n$ are similar to those results obtained from the tested ocean color models and datasets. In normal, lognormal, or exponential distributions, RMSE typically present uptrends (Fig. 4(a), 4(c), and 4(d)). MAE and RMSE with uniform error distribution present the similar flat trends (Fig. 4(b)). MAE with lognormal and exponential error distributions present slightly upward trends within the first few samples (Fig. 4(c) and 4(d)). For all error datasets, $U_A$ always presents a downtrend.

Fig. 4 shows only the results derived from the simulated error distributions with specific parameters. For example, the normal distribution shown in Fig. 4(a) is standard with parameters mean $\mu = 0$ and standard deviation $\sigma = 1$, and for the uniform distribution, errors are uniformly distributed between 0 and 1. If we set different error distribution parameters, the resultant trends of MAE, RMSE, and $U_A$ are similar to those shown in Fig. 4.

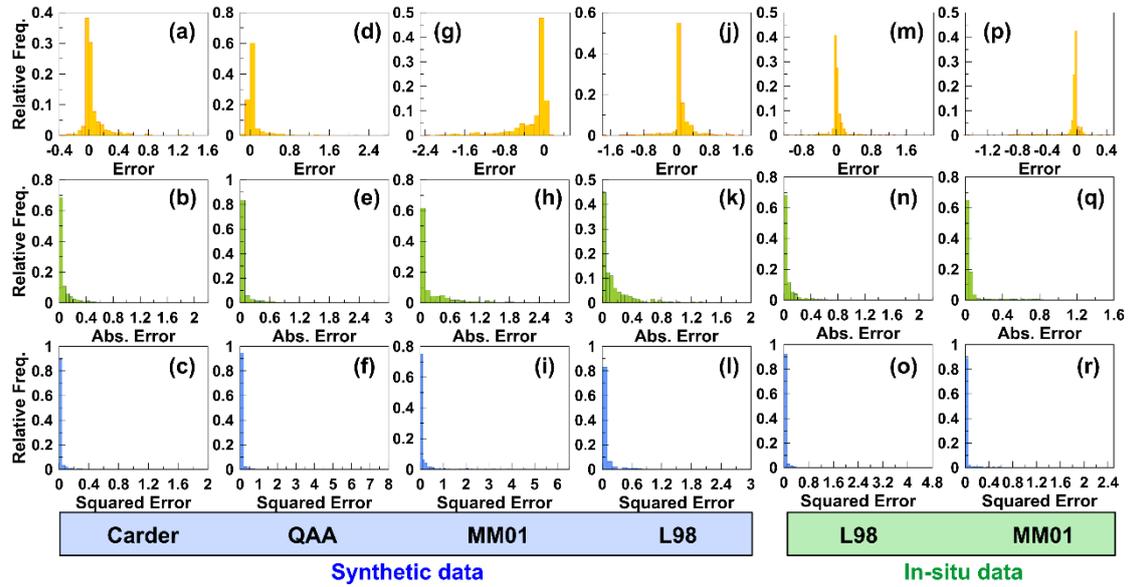

Figure 2. Error distributions of total absorption coefficients $a$(440) modeled from four ocean color algorithms and two datasets.

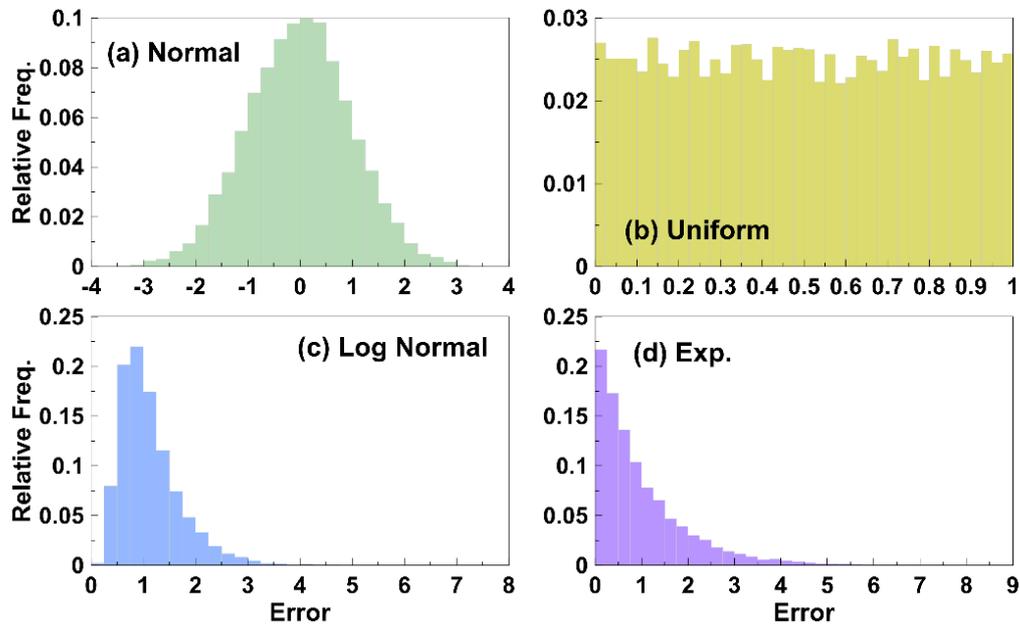

Figure 3. Histograms of computer simulated errors with four distributions: (a) normal, (b) exponential, (c) lognormal, and (d) uniform.

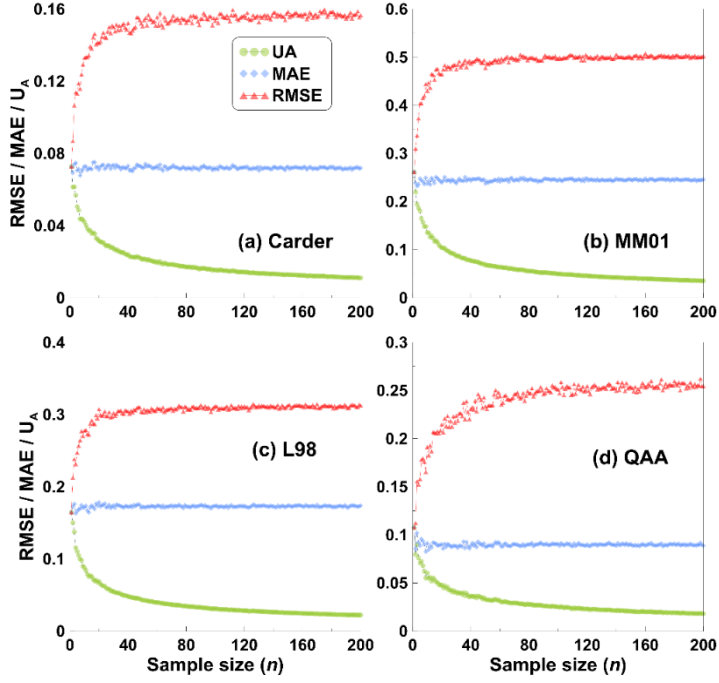

Figure 4. Statistical parameter trends vs. sample sizes calculated from simulated errors with four distributions: (a) normal, (b) exponential, (c) lognormal, and (d) uniform.

### 3.3 Theoretical statistical analysis

Suppose an error set $E = \{e_1, e_2, \ldots, e_n\}$ contains $n$ absolute errors and their mean is $e_{mean}$. $E_{max} = \{e_1^{max}, e_2^{max}, \ldots, e_n^{max}\}$ is the same set of $E$ but errors are reordered from the maximum to the minimum, that is, $e_1^{max}$ is the maximal error and $e_n^{max}$ is the minimal error in $E$. $E_{min} = \{e_1^{min}, e_2^{min}, \ldots, e_n^{min}\}$ is the reverse order of $E_{max}$. If we randomly select $i$ errors from $E$ forming an error subset $E_{s1}$ and repeat the selection $m$ times, and then we obtain $m$ subsets $E_{s1} = \{e_1^{s1}, e_2^{s1}, \ldots, e_i^{s1}\}$, $E_{s2} = \{e_1^{s2}, e_2^{s2}, \ldots, e_i^{s2}\}$, …, $E_{sm} = \{e_1^{sm}, e_2^{sm}, \ldots, e_i^{sm}\}$, and their respective MAEs are $MAE_{s1}$, $MAE_{s2}$, …, $MAE_{sm}$. We take the mean of these MAEs as the mean MAE of the $i$ errors (namely $i$ samples), denoted by $MAE_i$, that is,

$$MAE_i = \frac{\sum_{k=1}^{m} MAE_{sk}}{m} \qquad (5)$$

If $i = n$, then $E_{s1} = E_{s2} = \ldots = E_{sm} = E$, so $MAE_{s1} = MAE_{s2} = \ldots = MAE_{sm} = e_{mean}$, then from Eq. (5) it is easy to know that $MAE_n = e_{mean}$. If $i = 1$, any subset $E_{sm}$ only contains one error $e_1^{sm}$, then Eq. (5) turns to

$$MAE_1 = \frac{\sum_{k=1}^{m} e_1^{Sk}}{m} \qquad (6)$$

where $e_1^{Sk}$ could be any error in $E$. The equation could be rewritten to

$$MAE_1 = \frac{\sum_{k=1}^{n} p_k e_k}{m} \qquad (7)$$

where $p_k$ tells the times that $e_k$ was selected to be the $e_1^{Sk}$. Because there are $n$ errors but we randomly selected $m$ times, so each error should be selected $p_k = m/n$ times, i.e., the possibility of an error being selected. For example, if we have 500 samples and selected 1000 times, then each sample is likely to be selected twice. Therefore, if we made sufficient random selections, namely, $m$ is big enough, then Eq. (7) will turn to

$$MAE_1 = \frac{\sum_{k=1}^{n} \frac{m}{n} e_k}{m} = \frac{\sum_{k=1}^{n} e_k}{n} = e_{mean} \qquad (8)$$

The Eq. (8) indicates that MAEs at the begin point ($i = 1$, the smallest sample size) and the end point ($i = n$, the largest the sample size) of the trend line are both equal to $e_{mean}$. The below equation proves that at any point of the trend line, MAE is always approximately equal to $e_{mean}$.

$$MAE_i = \frac{\sum_{j=1}^{m} \sum_{k=1}^{i} e_k^{Sj}}{mi} = \frac{\sum_{k=1}^{n} \frac{mi}{n} e_k}{mi} = e_{mean} \qquad (9)$$

The above analysis shows that the mean MAE of a given sample size is expected to be the mean of the sample population. However, a single MAE of the given sample size should be fallen within some range, see $d_i$ in Fig. 5(a). The $d_i$ can be calculated by the below equation:

$$d_i = \frac{\sum_{m=1}^{i} e_m^{max} - \sum_{m=1}^{i} e_m^{min}}{i} \qquad (10)$$

Then we calculate $\Delta d$, the variation of $d$ between two adjacent sample sizes $i$ and $i + 1$, using the below equation:

$$\Delta d = d_{i+1} - d_i = \sum_{m=1}^{i} e_m^{min} - \sum_{m=1}^{i} e_m^{max} + i e_{i+1}^{max} - i e_{i+1}^{min} \qquad (11)$$

Because $e_i^{max} \geq e_{i+1}^{max}$ and $e_i^{min} \leq e_{i+1}^{min}$, then $\Delta d \leq 0$ (equals 0 only if $e_1^{max} = e_2^{max} = \cdots = e_{i+1}^{max}$ and $e_1^{min} = e_2^{min} = \cdots = e_{i+1}^{min}$), meaning that the range $d$ becomes narrower and narrower as $n$ becomes larger. It can be proved that the similar $d$ calculated from RMSE holds the same trend of becoming narrower. If we take $d$ as the uncertainties of the MAE or RMSE, then the larger $n$ making the smaller $d$ indicates that as we have more and more samples, the uncertainty of the model should be smaller and smaller until the MAE/RMSE of the sample population. MAE/RMSE itself, however, does not imply the trend clearly. Instead, it usually demonstrates a flat line or sometimes an uptrend demonstrated in Fig. (1) and (4).

Above we already proved that the flat trend of MAE is the ideal case if we take the mean value of $m$ MAEs calculated from $m$ selections and $m$ is big enough. Note that in this case, the flat trend of MAE does not depend on error distributions. Instead of using mean, if we take the median value of the $m$ MAEs as the MAE of the given sample size, then MAE may bear a slight uptrend or downtrend within the first few sample sizes, depending on error distributions. From Eq. 5(b) we know that if using median, then $MAE_n$ is still the mean of all errors, but $MAE_1$ should be the median rather than the mean of all errors. Therefore, for all errors, if their median is equal to their mean, the MAE will still be a flat trend, otherwise it will be an uptrend if their median is less than the mean, and a downtrend if their median is greater than their mean. We know that the median-mean relationship depends on error distributions – for example, whether errors are symmetrically distributed with respect to their mean.

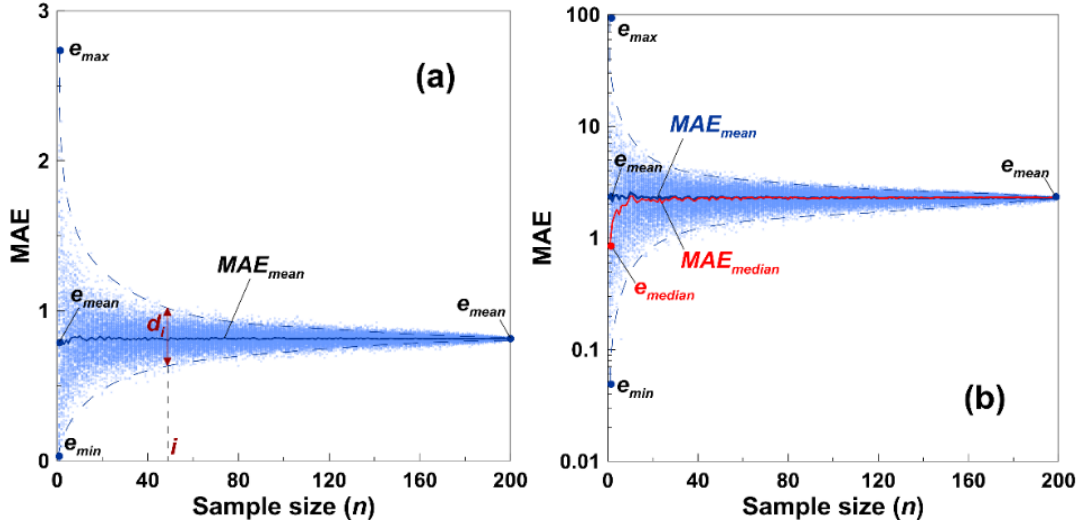

Figure 5. MAE distributions vs. sample sizes with (a) normal and (b) lognormal error distributions

Now let us consider error distributions of practical models. Based on the results of our ocean color remote sensing algorithm/dataset tests, model errors are typically with lognormal-shape or exponential-shape distributions, particularly for the absolute and squared errors, see histograms in Fig 2. The resultant distribution histograms are reasonable because most of model errors are expected to be small while large errors are not so frequent, provided that the model was properly built and validation data were properly collected. For such error distributions, the error median is usually less than the error mean (see results in Table 1), leading to an uptrend median MAE within the first few samples. Note that not only the median MAE, but also the mean MAE will likely to be an uptrend since the flat trend of mean MAE only occurs when the selection times $m$ is big enough. If only selecting, say, two or three samples, then their mean MAEs are very likely to be less than the $e_{mean}$, because the two or three samples are very likely to make small errors.

In the ideal case, if errors have a standard normal distribution with mean $\mu = 0$ (no bias or systematic errors), then the absolute and squared errors should be folded normal distribution and chi-squared distribution with degree of freedom 1, respectively. The two distributions are both median-less-than-mean distributions and hence their MAE and RMSE are both with the uptrends as those shown in our testing results.

Given MAE usually holds a flat or increasing trend, the RMSE and $U_A$ obtained from the same error set usually present more significant uptrend and downtrend, respectively. Their trends can be explained by the relations among MAE, RMSE, and $U_A$. It is known that

$$MAE \leq RMSE \leq \sqrt{n}MAE \qquad (12)$$

Because $U_A = \frac{RMSE}{\sqrt{n}}$, then the upper and lower limits of $U_A$ are

$$\frac{MAE}{\sqrt{n}} \leq U_A \leq MAE \qquad (13)$$

If we assume MAE being an ideal constant (a perfect flat trend), and plot the functions $f_{MAE}(n) = MAE$, $f_{RMSE}(n) = \sqrt{n}MAE$ and $f_{U_A}(n) = \frac{MAE}{\sqrt{n}}$ in Fig. 6, where the three curves all begin at the same point $n = 1$ with $MAE_1 = RMAE_1 = U_{A1}$. Then there are a RMSE zone above the MAE line and a $U_A$ zone below the MAE line. The RMSE and UA trend lines should be within the two zones, implying there are likely to be an uptrend and a downtrend, respectively, at least for the first few samples. If MAE bears the first uptrend, see Fig. 6(c), then RMSE always presents a similar uptrend while UA is possible to show a downtrend ($U_{A1}$ in Fig. 6(c)), or a similar uptrend for the first a few samples and then turn to a downtrend because of the division by $\sqrt{n}$ ($U_{A2}$ in Fig. 6(c)). If MAE bears the first downtrend, see Fig. 6(b), then $U_A$ always presents a similar downtrend while RMSE is possible to show a similar downtrend ($RMSE_1$ in Fig. 6(b)), or still an uptrend for all samples due to error distributions and effect of multiplication by $\sqrt{n}$ ($RMSE_2$ in Fig. 6(b)). Note that the case of downward MAE only occurs when most errors are significantly large, but small errors make up a small proportion of all errors. This case, however, does not frequently happen in remote sensing modelling and validation. If the case happened, then it should tell that the model has not been well made or the sample data is not appropriate for evaluating the model. For instance, if we use samples collected from turbid inland waters to assess a chlorophyll model based on clear seawater, then its errors may be the case. Typically, the most common cases are those MAE, RMSE and $U_{A2}$ in Fig. 6(c), that is, MAE presents a flat or slightly upward trend, RMSE presents a relatively significant uptrend, and meanwhile $U_A$ presents a significant downtrend. And indeed, these cases have been verified by our practical algorithm tests and simulated error tests.

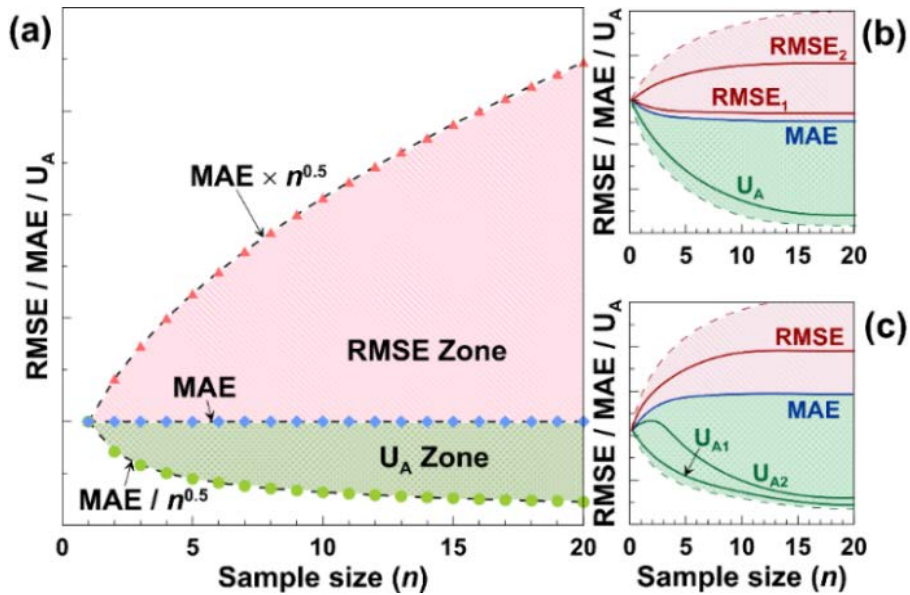

Figure 6. The possible zones of RMSE and UA with respect to a given MAE trend: (a) flat, (b) downward, and (c) upward.

**3.4 Discussions**

The concepts of uncertainty and error are easy to be muddled with each other. We would like to emphasize that model's uncertainty is different from its error. Uncertainty indicates the reliability of a model while error indicates the accuracy of the model. To evaluate model performance, it is necessary to not only evaluate model's error, but also evaluate its reliability. Small errors (small MAE and RMSE) do not certainly imply small uncertainty, because if the small errors were only based on a few of samples, we are unable to rely on the model. Given the remotely sensed objects are usually quite complicated, small sample sizes are unlikely to represent all possible aspects and features of these objects.

Once a model was established, we would expect its inherent accuracy to be independent of the validation samples, that is, the model would not be more accurate only because of more samples being used to evaluate the model. In this perspective, MAE is a better parameter for accuracy evaluation than RMSE because MAE generally does not vary significantly with sample size. The uptrend of RMSE within the first 20 ~ 50 samples is a misleading drawback, because we remote sensing modellers are not encouraged to collect more samples: the more samples we take, the larger the RMSE will be, and the less accurate the model will be. In comparison, $U_A$ is a good parameter which correctly reflects the importance of sample size in modelling and validation. It can correct the misleading evaluation by only using RMSE. Remote sensing modelers are then encouraged to collect more samples – although their models may have relatively large RMSE due to large sample sizes, the model uncertainties (reliabilities) are tended to be lower than those of models based on small sample size.

Despite its drawbacks, RMSE is calculated by using the sum of squared errors, compared to MAE, so it also bears some advantages: (1) the sum of squared errors avoids the absolute value used in MAE, which is highly undesirable in many mathematical calculations [4], (2) in data assimilation field, the sum of squared errors is often defined as the cost function to be minimized by adjusting model parameters, while MAEs are definitely not preferred over RMSEs [4], and (3) the analytically based minimization of a sum-of-squares-based measure during fitting is less cumbersome [9]. The above advantages of RMSE are also applied to $U_A$ since $U_A$ uses the same sum-of-squares-based formula.

$U_A$ is not only an indicator of model uncertainty, but also an indicator of the uncertainty of MAE or RMSE, namely, the range $d_i$ in Fig. 5(a). We are unable to calculate the accurate $d_i$ since the sample population is usually unable to obtain. Because $U_A$ holds a similar downtrend to $d_i$, it can be used as a proxy of $d_i$, meaning the more samples we have, the less uncertainty (the smaller $d_i$) we know about the MAE or RMSE of the sample population.

It is necessary to mention that although errors are combined into the calculation of $U_A$, we cannot use $U_A$ to indicate model errors, because it may arise the situation that MAE and RMSE are very large, but $U_A$ is very small due to an extremely large sample size. In order to make the assessment of model performance more comprehensive, our suggestion is reporting model accuracy (using MAE and RMSE) and uncertainty (using $U_A$), as well as other useful statistical parameters if necessary.

## 4. Conclusion

Tests of simulated error datasets and in-situ/synthetic ocean color remote sensing datasets and models show that RMSE, MAE, and $U_A$ all vary with the number of observations *n* but present different trends. Typically, RMSE presents an upward trend, MAE in most cases presents a flat trend but sometimes also a slightly uptrend, while $U_A$ always bears a smooth downward trend. The theoretical analyses of these statistical parameters elucidate why there are such trends. The uptrends of RMSE indicate that when sample size is small, model/algorithm performances are likely to be overestimated if we only use RMSE as evaluation parameters. The flat trends of MAE and the same trends of RMSE for large sample size imply that model performances are no longer affected by sample size so that model reliability associated with lager sample sizes has not been well expressed.

Not only RMSE, MAE and other parameters are needed to evaluate the accuracy of a model, but also we need to use $U_A$ to assess its reliability. $U_A$ not only inherits the advantages of RMSE, but also presents a decreasing trend with the increasing of sample size *n*, which is consistent with the similar trends of MAE/RMSE that their uncertainties $d_i$ become lower and lower as the sample size goes to the population. It should be emphasized that the model error is not equal to model uncertainty. Given the complexity of the modelled environmental objects, only evaluating model accuracy without uncertainty may make a misleading assessment of model performance.

**Acknowledgement**
This research is supported the National Natural Science Foundation of China (No. 41471346) and Natural Science Foundation of Zhejiang Province (No. LY17D010005).